\documentclass{emulateapj}

\usepackage{bm}

\newcommand{\beq}{\begin{equation}}
\newcommand{\eeq}{\end{equation}}
\newcommand{\bea}{\begin{eqnarray}}
\newcommand{\eea}{\end{eqnarray}}
\newcommand{\req}[1]{equation (\ref{#1})}
\newcommand{\dd}{\mathrm{d}}
\newcommand{\alphaf}{\alpha_\mathrm{f}}
\newcommand{\gcc}{\mbox{g cm$^{-3}$}}
\newcommand{\kB}{k_\mathrm{B}}
\newcommand{\mel}{m_e}

\newcommand{\Msun}{M_\odot}

\newcommand{\bP}{\bm{P}}
\newcommand{\Bq}{B_{\rm Q}}
\newcommand{\chiH}{\chi^\mathrm{H}}
\newcommand{\chiA}{\chi^\mathrm{A}}
\newcommand{\ChiH}{\bm{\chi}^\mathrm{H}}
\newcommand{\ChiA}{\bm{\chi}^\mathrm{A}}
\newcommand{\eX}{\bm{e}^\mathrm{X}}
\newcommand{\eO}{\bm{e}^\mathrm{O}}
\newcommand{\mH}{m_\mathrm{H}}
\newcommand{\nueff}{\nu_\mathrm{eff}}
\newcommand{\omc}{\omega_{\mathrm{c}e}} 
\newcommand{\omci}{\omega_{\mathrm{ci}}} 
\newcommand{\ompl}{\omega_\mathrm{pl}} 
\newcommand{\Teff}{T_\mathrm{eff}}
\newcommand{\xH}{x_\mathrm{H}}

\shorttitle{Polarization in Magnetic Neutron Star Atmospheres}

\begin{document}

\title{Electromagnetic Polarization in Partially Ionized 
Plasmas with Strong Magnetic Fields
and Neutron Star Atmosphere Models}

\author{Alexander Y. Potekhin,\altaffilmark{1,2}
Dong Lai,\altaffilmark{3}
Gilles Chabrier,\altaffilmark{4}
and
Wynn C. G. Ho\altaffilmark{5,6}}
\altaffiltext{1}{Ioffe Physico-Technical Institute,
  Politekhnicheskaya 26, 194021 St.~Petersburg, Russia;
{palex@astro.ioffe.ru}}
\altaffiltext{2}{Isaac Newton Institute of Chile, 
         St.~Petersburg Branch, Russia}
\altaffiltext{3}{Center for Radiophysics and Space Research, 
 Department of Astronomy,
 Cornell University,
 Ithaca, NY 14853;
{dong@astro.cornell.edu}}
\altaffiltext{4}{Ecole Normale Sup\'erieure de Lyon
  (CRAL, UMR CNRS No.\ 5574), 
46 all\'ee d'Italie, 69364 Lyon Cedex 07, France;
{chabrier@ens-lyon.fr}}
\altaffiltext{5}{Kavli Institute for Particle Astrophysics and Cosmology,
Stanford University, PO Box 20450, Mail Stop 29, Stanford, CA 94309;
{wynnho@slac.stanford.edu}}
\altaffiltext{6}{Hubble Fellow}

\begin{abstract}

Polarizability tensor of a strongly magnetized plasma and the
polarization vectors and opacities of normal electromagnetic waves are
studied for the conditions typical of neutron star atmospheres, taking
account of partial ionization effects. Vacuum polarization is also 
included using a new set of fitting formulae that are accurate for
wide range of field strengths. The full account of the coupling of the
quantum mechanical structure of the atoms to their center-of-mass
motion across the magnetic field is shown to be crucial for the correct
evaluation of the polarization properties and opacities of the plasma.
The self-consistent treatment of the polarizability and absorption
coefficients proves to be necessary if the ionization degree of the
plasma is low, which can occur in the atmospheres of cool or
ultramagnetized neutron stars. 
Atmosphere models and spectra are presented
to illustrate the importance of such self-consistent treatment.

\end{abstract}


\maketitle


\section{Introduction}
\label{sect:intro}

In recent years, thermal or thermal-like radiation has been detected
from several classes of isolated neutron stars (NSs): radio pulsars
with typical magnetic fields $B\sim10^{12}$--$10^{13}$ G, ``dim'' NSs
whose magnetic fields are mostly unknown, anomalous X-ray pulsars and
soft gamma-ray repeaters with $B$ possibly $\sim10^{14}$--$10^{15}$ G 
\citep*[see, e.g.,][
for reviews]{BeckerAschenbach,Haberl-COSPAR,IMS02,PZ03}. The spectrum of
thermal radiation is formed in a thin atmospheric layer (with scale
height $\sim0.1$--$10$ cm and density $\rho\sim10^{-2}$--$10^3$ \gcc)
that covers the stellar surface. Therefore, a proper interpretation of
the observations of NS surface emission requires understanding of
radiative properties of these magnetized atmospheres. 

\citet*{Shib92} \citep*[see also][]{SZ95,Pavlov95} presented the first 
model of the NS atmospheres with strong magnetic fields,
assuming full ionization. Variants of this model were constructed by
\citet*{Zane00,Zane01,HoLai,HoLai02,Ozel01,Lloyd03}. 
An inaccurate treatment of
the absorption due to free-free transitions in strong magnetic fields
in the earlier models \citep{Pavlov95} has been corrected by
\citet{PC03}; this correction has been taken into account in later
models \citep{hoetal03,Ho-COSPAR,Lloyd03}. 
Recent work \citep{HoLai02,LaiHo02,LaiHo03a} has shown that in 
the magnetar field regime ($B\gtrsim 10^{14}$~G) vacuum polarization 
significantly affects the emergent spectrum from the atmosphere;
for weaker fields, vacuum polarization can still leave an unique
imprint on the X-ray polarization signals \citep{LaiHo03b}.

Because the strong magnetic field significantly increases the binding
energies of atoms, molecules, and other bound states \citep[see][
for a review]{Lai01}, these bound states may have abundances
appreciable enough to contribute to the opacity in the atmosphere. For
calculation of this contribution, the non-trivial coupling of the
center-of-mass (CM) motion of the atom to its internal structure 
\citep[e.g.,][ and references therein]{P94} can be important. 
Also, because of the relatively high
atmosphere density, a proper treatment should take account of
 the plasma nonideality
that leads to Stark broadening 
and pressure ionization. Recently,
thermodynamically consistent equation of state (EOS) 
and opacities have been obtained for a magnetized,
partially ionized H plasma for $8\times10^{11}$ G $\la B\leq10^{15}$ G
\citep*{PCS99,PC03,PC04}.
These EOS and opacities have been implemented by
\citet{hoetal03,Ho-COSPAR} for modeling NS atmospheres. For
the typical field strengths $B=10^{12}$--$10^{13}$~G this modeling
showed that, although the spectral features due to neutral atoms lie at
extreme UV and very soft X-ray energy bands and therefore are difficult
to observe, the continuum flux is also different from the fully ionized
case, especially at lower energies, which can affect fitting of the
observed spectra. For the superstrong field $B\gtrsim 10^{14}$~G,
\citet{hoetal03} showed that the vacuum polarization effect not only
suppresses the proton cyclotron line, but also suppresses spectral
features due to bound species.

It is well known \citep[e.g.,][]{Ginzburg,Mesz} that under typical
conditions (e.g., far from the resonances) radiation propagates in a
magnetized plasma in the form of two so-called extraordinary and
ordinary normal modes. The polarization vectors of these modes,
$\eX$ and $\eO$ are determined by the
Hermitian part ($\ChiH$) of the complex polarizability
tensor ($\bm{\chi}$) of the plasma. Our previous treatment of these modes
in partially ionized atmospheres \citep{PC03,PC04,hoetal03,Ho-COSPAR}
was not quite self-consistent, because the effect of the presence of
the bound states on the polarization of normal modes was neglected: we
adopted the same polarization vectors as in the fully ionized plasma,
assuming \citep{PC03} that the effect of bound states on these vectors
should be small provided the ionization degree of the plasma is high.
However, this hypothesis (related to $\ChiH$) may be
called into question, based on the observation that the absorption
coefficients (corresponding to the {anti-Hermitian} part of the
complex polarizability tensor, $i\ChiA$) are strikingly affected
by the presence of even a few percent of the atoms.

In this paper, we study the polarizability tensor, the polarization
vectors of the normal waves, and the opacities of the partially ionized
nonideal hydrogen plasma in strong magnetic fields in a self-consistent
manner, using the technique applied previously by \citet{BulikPavlov}
to the case of a monatomic ideal hydrogen gas. In \S\ref{sect:general}
we introduce basic definitions and formulae to be used for calculation
of the plasma polarizability (new fitting formulae for the vacuum
polarizability are given in the Appendix). An approximate model based
on a perturbation theory, which explains the importance of the CM
coupling for plasma polarizability, is described in \S\ref{sect:pert}.
The results of numerical calculations of the plasma polarizability are
presented in \S\ref{sect:chi-real}, and the consequences for the
polarization and opacities of the normal modes are discussed in
\S\ref{sect:res}. In \S\ref{sect:spectra} we present examples of NS
thermal spectra, calculated using the new opacities, compared with the
earlier results. In \S\ref{sect:concl} 
we summarize our results, outline the range of their
applicability, and discuss unsolved problems.

\section{General Formulae for Polarization in a Magnetized Plasma}
\label{sect:general}

\subsection{Complex Polarizability Tensor}

The propagation of electromagnetic waves in a medium is
described by the wave equation that is obtained 
from the Maxwell equations involving the tensors 
of electric permittivity $\bm{\epsilon}$,
magnetic permeability $\bm{\mu}$, 
and electrical conductivity $\bm{\sigma}$.
It is convenient \citep[e.g.,][]{Ginzburg} to introduce 
the complex dielectric tensor 
$\bm{\epsilon}'=\bm{\epsilon}+4\pi i \bm{\sigma}/\omega$.
In the strong magnetic field, 
it should include the vacuum polarization. 
When the vacuum polarization is small, it can be linearly added
to the plasma polarization. Then we can write
\beq
 \bm{\epsilon}' = \mathbf{I} + 4\pi\bm{\chi} + 4\pi\bm{\chi}^\mathrm{vac},
\label{linear}
\eeq
where $\mathbf{I}$ is the unit tensor, 
$\bm{\chi}=\ChiH+i\ChiA$ is the complex polarizability tensor of the plasma,
and $\bm{\chi}^\mathrm{vac}$ is the polarizability tensor of the vacuum.

In the Cartesian coordinate system $x'y'z'$ with 
unit vectors $\hat\mathbf{x}'$, $\hat\mathbf{y}'$, $\hat\mathbf{z}'$,
where $\hat\mathbf{z}'$ is along magnetic field 
$\bm{B}$, the electric permittivity tensor 
of a plasma in the dipole approximation, 
the dielectric vacuum correction,
and the inverse magnetic permeability 
of the vacuum can be written, respectively, as
\citep[e.g.,][ and references therein]{HoLai02}
\bea
&&
 \mathbf{I} + 4\pi {\bm\chi} = 
 \left[ \begin{array}{ccc}
 \varepsilon & ig & 0 \\
 -ig & \varepsilon & 0 \\
 0 & 0 & \eta 
 \end{array} \right],
\label{eps-p}
\\&&
 4\pi \bm{\chi}^\mathrm{vac} = 
 \mathrm{diag}(\hat{a}, \hat{a}, \hat{a}+q ),
\label{chivac}
\\&&
 \bm{\mu}^{-1} = \mathbf{I} + \mathrm{diag}(\hat{a}, \hat{a}, \hat{a}+m),
\eea
where $\hat{a}$, $q$, and $m$ are vacuum polarization coefficients
which vanish at $B=0$. The formulae for calculation of these 
coefficients are given in the Appendix. 
Quantities $\varepsilon$, $\eta$, and $g$ are well known 
for fully ionized ideal plasmas \citep[e.g.,][]{Ginzburg}.
In this paper we shall calculate $\bm\chi$
for partially ionized hydrogen plasmas
at typical NS atmosphere conditions.

The complex polarizability tensor of a plasma becomes diagonal, 
$\bm{\chi}=\mathrm{diag}(\chi_{+1},\chi_{-1},\chi_0)$ in the 
cyclic or rotating coordinates, where the cyclic unit vectors are
defined as $\hat\mathbf{e}_0 = \hat\mathbf{z}'$, $\hat\mathbf{e}_{\pm1} =
(\hat\mathbf{x}' \pm i \hat\mathbf{y}')/\sqrt{2}$. 
The real parts of
the components $\chi_\alpha$ ($\alpha=\pm1,0$) determine the Hermitian
tensor $\ChiH$, which describes the refraction and
polarization of waves, and their imaginary parts determine the
anti-Hermitian tensor $i\ChiA$, responsible for the
absorption. According to \req{eps-p}, 
\beq
4\pi\chi_{\pm1}=\varepsilon - 1
\pm g,\quad 4\pi\chi_0 = \eta - 1.
\eeq 
Note that in the cyclic representation, 
the general symmetry relations for the polarizability tensor 
take the form
\beq
  \chiA_\alpha(-\omega)=-\chiA_{-\alpha}(\omega),
\quad
  \chiH_\alpha(-\omega)=\chiH_{-\alpha}(\omega) .
\label{sym}
\eeq

\subsection{Relation Between the Plasma 
Polarizability and Absorption}

General expressions in the
dipole approximation for $\chiA_\alpha$ and $\chiH_\alpha$ through 
frequencies and oscillator strengths of quantum transitions 
in a magnetized medium are given, e.g., by \citet{BulikPavlov}.
For transitions between two 
stationary quantum states $i$ and $f$
with energies $E_i$ and $E_f=E_i+\hbar\omega_{fi} > E_i$
and number densities of the occupied states $n_i$ and $n_f$,
the absorption coefficient for the basic polarization $\alpha$
equals \citep[see, e.g.,][]{Armstrong}
\beq
  \mu^{if}_\alpha \equiv \rho\hat\kappa_\alpha(\omega)
  = \frac{2\pi^2 e^2}{\mel c}\,(n_i-n_f)\,f^{if}_\alpha
   \delta(\omega-\omega_{fi}),
\label{mu-delta}
\eeq
where 
$f^{if}_\alpha = 
(2\mel\omega_{fi}/\hbar)\,|\langle f|\bm{r}|i\rangle
          \cdot\hat\mathbf{e}_\alpha|^2$
is the oscillator strength for the transition $i\rightarrow f$.
These partial absorption coefficients sum up into the total
$\mu_\alpha = \sum_{i,f(E_f>E_i)} \mu^{if}_\alpha$,
where $\sum_{i,f}$ includes integration over continuum states.
Then the equation
\beq
  \chiA_\alpha(\omega) = \frac{c}{4\pi\omega}\, \mu_\alpha(\omega),
\label{chi-mu}
\eeq
together with the first symmetry relation (\ref{sym})
yield
\bea
  \chiA_\alpha(\omega) &=& \frac{\pi e^2}{2\mel\omega}\,\sum_{i,f (E_f>E_i)}
  (n_i-n_f) 
 \nonumber\\&&\times
  \left[ f^{if}_\alpha \,\delta(\omega-\omega_{fi})
   + f^{if}_{-\alpha} \,\delta(\omega+\omega_{fi}) \right].
\label{chiA-general}
\eea

Once $\chiA$ is known, $\chiH$ can be calculated
from the Kramers-Kronig relation \citep[ \S123]{LaLi-Stat}
or its modification \citep{BulikPavlov},
\beq
  \chiH_\alpha(\omega) = \frac{1}{\pi\omega}\,\mathcal{P}
   \int_{-\infty}^\infty \frac{\omega'\chiA_\alpha(\omega')
   }{\omega'-\omega}\,\dd\omega' ,
\label{KK}
\eeq
where $\mathcal{P}\!\int$ means the principal value of the integral.
From equations (\ref{chiA-general}) and (\ref{KK}),
\beq
  \chiH_\alpha(\omega) = - \frac{e^2}{2\,\mel\omega}
\!\!\!\!\!\!  \sum_{i,f\, (E_f>E_i)\rule{0pt}{2ex}} \!\!\!\!\!\!
  (n_i-n_f) \left[ \frac{f^{if}_\alpha}{\omega-\omega_{fi}}
   + \frac{f^{if}_{-\alpha}}{\omega+\omega_{fi}} \right] .
\label{chiH-general}
\eeq
Taking into account equations (\ref{chi-mu}) and (\ref{sym}),
we can present relation (\ref{KK}) in the form convenient for calculation
at $\omega>0$:
\bea
  \chiH_\alpha(\omega) &=& \frac{c}{4\pi^2\omega}\, \bigg\{
    \int_0^\omega \big[\,\mu_\alpha (\omega+\omega')
     - \mu_\alpha (\omega-\omega')\,\big]
     \frac{\dd\omega'}{\omega'}
\nonumber\\&&
   + \int_{2\omega}^\infty \frac{\mu_\alpha(\omega')}{\omega'-\omega}
     \,\dd\omega'
   - \int_0^\infty \frac{\mu_{-\alpha}(\omega')}{\omega'+\omega}
     \,\dd\omega' \bigg\} .
\label{KK-mu}
\eea

If we replace delta-function in \req{mu-delta}
by the Lorentz profile 
$(\nu^{if}_\alpha/\pi)\,
[\,(\omega-\omega_\alpha)^2+(\nu^{if}_\alpha)^2\,]^{-1}$,
where $\nu^{if}_\alpha$ is a damping rate for a given transition,
then 
equations (\ref{chiA-general}) and (\ref{chiH-general}) 
can be combined to give
\begin{eqnarray}
  \chi_\alpha(\omega) &=& - \frac{e^2}{2\,\mel\omega}
  \sum_{i,f\, (E_f>E_i)}
  (n_i-n_f) \Bigg[ 
  \frac{f^{if}_\alpha}{\omega-\omega_{fi}+i\nu^{if}_\alpha}
 \nonumber\\&&
   + \frac{f^{if}_{-\alpha}}{\omega+\omega_{fi}+i\nu^{if}_{-\alpha}}
   \Bigg].
\label{chi-general}
\end{eqnarray}

\section{Analytic Model of Polarization 
of Atomic Gas: Effect of Center-of-Mass Motion}
\label{sect:pert}

As mentioned in \S\ref{sect:intro}, in strong magnetic fields the
internal structure of an atom is strongly coupled to its CM motion
perpendicular to the field. This coupling (referred to as ``CM
coupling'') has a significant effect on the radiative opacities and
dielectric property of the medium. Before presenting numerical results
for the polarizability tensor of a partially ionized plasma in
\S\ref{sect:chi-real}, it is useful to consider an analytic model to
illustrate this CM coupling effect.

Consider an atomic gas in which the atom possesses only two energy
levels, with the upper level having the radiative (Lorentz) width $\nu$,
assumed to be constant for simplicity.
The energies of the (moving) atom in the ground state and the excited
state are denoted by $E_1(\bP)$ and $E_2(\bP')$ respectively, where $\bP$ 
and $\bP'$ are the CM pseudo-momenta, and the subscripts 1 and 2 specify 
the internal degree of freedom of the atom. [In strong magnetic fields,
the internal quantum numbers are $(s,\nu_z)$, where $s=0,1,2,\cdots$ 
measures the relative angular momentum between the proton and electron, 
and $\nu_z$ is related to the number of nodes in the $z$ direction.]
If there were no CM coupling, we would have 
$E_{1,2}(\bP)=E_{1,2}(0)+
\bP^2/(2\mH)$ and 
$\omega_{21}\equiv(E_2-E_1)/\hbar=$constant. 
Therefore, in this case
equation (\ref{chi-general}) would yield
\beq
\chi_\alpha(\omega)\simeq -\,{e^2n_1\over 2m_e\omega}
\left[{f_\alpha^{12}\over \omega-\omega_{21}+i\nu}
+ {f_{-\alpha}^{12}\over \omega+\omega_{21}+i\nu}\right].
\label{eq-wrong}
\eeq
This
would imply that even for very small neutral atom fraction,
the bound-bound transition can severely affect the dielectric property 
of the gas in the neighborhood of $\omega=\omega_{21}$
(for $\nu\ll\omega_{21}$).

However, in strong magnetic fields, the energy associated with the transverse
CM motion of the atom cannot be separated from the internal energy.
We therefore have 
\begin{eqnarray}
\chi_\alpha(\omega) &=& -{e^2 n_1\over 2m_e\omega}\int\!
{p_1(P_\perp)\,\dd^2 P_\perp\over (2\pi\hbar)^2}
\bigg[
\frac{f_\alpha^{12}(\bP_\perp)}{\omega-\omega_{21}(P_\perp)+i\nu}
 \nonumber\\&&
+\frac{f_{-\alpha}^{12}(\bP_\perp)}{\omega+\omega_{21}(P_\perp)+i\nu}
\bigg] ,
\end{eqnarray}
where $\bP_\perp$ is the transverse pseudo-momentum, 
$p_1(P_\perp)\,\dd^2 P_\perp$ is the probability to find an atom
in the initial state ``1'' in an element $\dd^2 P_\perp$ near $\bP_\perp$
\citep[see][]{PCS99},
and we have 
used the fact that the oscillator strength is nonzero only when
$\bP=\bP'$ (in the dipole approximation). 
The CM coupling effect will significantly smooth out the
divergent behavior at $\omega=\omega_{21}$ in \req{eq-wrong}
(for $\nu\rightarrow 0$). 
This effect can be taken into account
using the perturbation result of \citet{PM93}, as was done
by \citet{BulikPavlov}, or numerically, using the techniques
of \citet{PP95} and \cite{PP97}.

In the perturbation theory, valid for small $P_\perp$,
the coupling reveals itself as an effective ``transverse'' mass 
$M_\perp > \mH$
acquired by the atom, which is different for different quantum levels. 
For the two-level atom,
the perturbation theory gives
$E_j(P_\perp)=E_j(0)+P_\perp^2/(2M_{\perp j})$ ($j=1,2$), and we find
\begin{eqnarray}
\chi_\alpha &\simeq& -\frac{e^2 n_1}{2\mel\omega}
\int_0^\infty\! \dd y \bigg[
\frac{f_\alpha^{12}\,e^{-y}}{\omega-\omega_{21}(0)+i\nu+a(T)\,y}
 \nonumber\\&&
+
\frac{f_{-\alpha}^{12}\,e^{-y}}{\omega+\omega_{21}(0)+i\nu+a(T)\,y}
\bigg] ,
\end{eqnarray}
where
$a(T)=
(\kB T/\hbar)(1-M_{\perp 1}/M_{\perp 2})$,
and we have used $p_1(P_\perp)\propto 
\exp[-P_\perp^2/(2M_{\perp 1} \kB T)]$.
At $\nu\to0$, only weak logarithmic divergence is present near 
$\omega=\omega_{21}$, compared to the much stronger 
$(\omega-\omega_{21})^{-1}$ divergence in \req{eq-wrong}.
Additional line broadening due to collisions 
(which implies $\nu\neq0$),
thermal
effect and pressure effect, not treated in this simple model, will 
further smooth out the ``divergent'' feature in the polarizability tensor. 

This simple model shows that a proper treatment of the CM coupling 
is important in calculating the polarizability tensor of a partially
ionized plasma. In practice, since the absorption coefficient 
$\mu_\alpha$ has already been calculated by \citet{PC03,PC04}, 
we find it is more convenient to apply the Kramers-Kronig
transformation (eq.~[\ref{KK-mu}]) to obtain $\chiH_\alpha$
(see \S\ref{sect:chi-real}). 

\section{Effect of Partial Ionization on Polarizability of a Strongly
Magnetized Hydrogen Plasma}
\label{sect:chi-real}

\begin{figure}\epsscale{1.}
\plotone{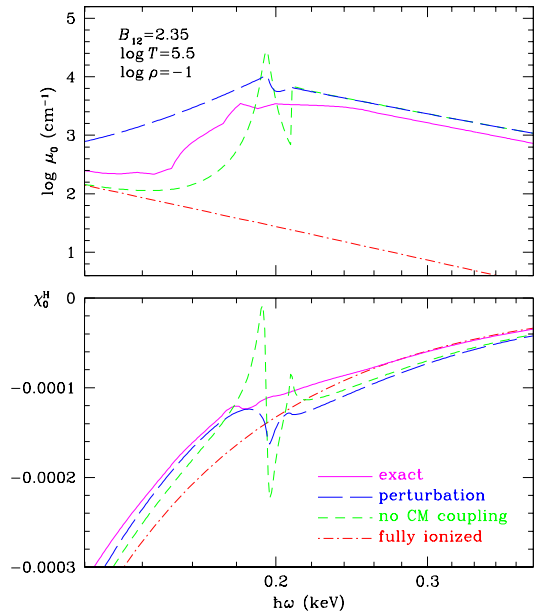}
\caption{Absorption coefficient (upper panel) and polarizability
(lower panel) of a hydrogen plasma 
for longitudinal polarization ($\alpha=0$) at $B=2.35\times10^{12}$~G,
$\rho=0.1$ \gcc, and $T=3.16\times10^5$~K, according to four
different models: fully ionized plasma (dot-dashed lines),
partially ionized plasma without the effect of coupling 
between CM motion and internal structure of the atom (short dashes),
with the magnetic broadening taken into account by perturbation
theory (long dashes) and numerical calculation 
beyond the perturbation approximation (solid lines).
\label{fig:kkpert0}}
\end{figure}

Some approximations for calculation of 
the complex dielectric tensor of a plasma have been discussed,
for example, by \citet{Ginzburg}. In particular, 
the ``elementary theory'' gives the expression widely used in the
past to describe polarization properties 
of the fully ionized electron-ion plasma 
\citep[e.g.,][]{Shib92,Zane00,HoLai,HoLai02}:
\beq
 4\pi\chi_\alpha = - \frac{\ompl^2
 }{
 (\omega+\alpha\omc) (\omega - \alpha\omci) + i \omega \nueff
 },
\label{chi-element}
\eeq
where $\omc=eB/\mel c$ and $\omci=ZeB/m_\mathrm{i} c$ are
the electron and ion cyclotron frequencies, 
$\ompl = (4\pi n_e e^2 / \mel)^{1/2}$ is the electron plasma frequency,
$\nueff$ is the effective damping frequency,
$m_\mathrm{i}$ is the ion mass,
$Ze$ is the ion charge,
and $n_e$ is the electron number density.
For the hydrogen plasma, the corresponding energies are
$\hbar\omc = 11.577\,B_{12}$ keV, $\hbar\omci = 6.3049\,B_{12}$ eV, 
where $B_{12}=B/10^{12}$~G,
and $\hbar\ompl= 0.0287\sqrt{\rho}$ keV, where $\rho$ is in \gcc.
In general, the damping frequency $\nueff$ depends on polarization and
photon frequency. Derivation of \req{chi-element} is 
based on the assumption that the electrons and ions lose their ordered
velocities (imparted by the electric field of the electromagnetic wave)
at the rate $\nueff$ which does not depend on the velocity. 
A more rigorous kinetic theory
gives $\chiA_\alpha$ and $\chiH_\alpha$ which cannot be in general described
by \req{chi-element}
using the same $\nueff$ \citep[e.g.,][ \S6]{Ginzburg}.
For instance, if we describe $\chiA_\alpha$ using
\req{chi-element} with some frequency-dependent $\nueff(\omega)$,
then the Kramers-Kronig transformation will give
$\chiH_\alpha$ which, in general, does not coincide with the real
part of \req{chi-element} with the same $\nueff(\omega)$,
although, for realistic $\nueff(\omega)$,
the difference should be small at $\omega\gg\ompl$.

Realistic models of the absorption coefficients of a partially ionized
plasma do not allow us to perform the Kramers-Kronig transformation
analytically. We evaluated the integrals in \req{KK-mu} numerically.
Since the integrand can be sharply peaked near resonance frequencies,
we employed the adaptive-stepsize Runge-Kutta integration
\citep[ \S16.2]{NRF}. The accuracy of the numerical transformation
was tested with the models that allow one to perform 
the Kramers-Kronig transformation analytically --
those given by equations (\ref{eq-wrong}) and (\ref{chi-element}) 
above, and by equations (49)--(53) of \citet{BulikPavlov}
-- and proved to be within 0.1\%.

\begin{figure}\epsscale{1.}
\plotone{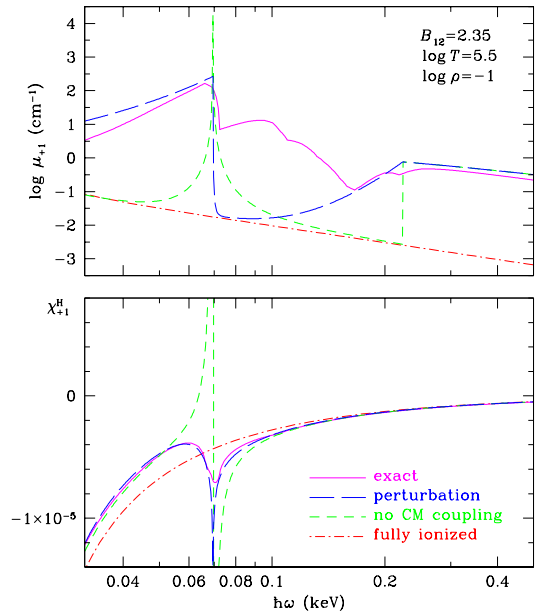}
\caption{Same as in Figure~\protect\ref{fig:kkpert0},
but for $\alpha=+1$.
\label{fig:kkpert1}}
\end{figure}

Let us first consider a model that neglects the CM coupling and the
plasma nonideality. The absorption coefficient of the fully ionized
component of the plasma includes contributions due to the free-free
absorption and scattering on free electrons an protons. For the atoms,
we include in consideration the bound-bound transition with the
largest oscillator strength (for every polarization) and the bound-free
transitions. In this case the absorption by the atom can be described
by analytic formulae. The bound-bound absorption cross section is
described by the Lorentz profile, where the effective damping width is
mainly contributed by the electron impact broadening \citep{PP95} and
can be evaluated using a fitting formula \citep{P98}. The bound-free
cross sections are well described in the adiabatic approximation 
(however this is not true with the CM coupling, see \citealt{PP97});
they are reproduced by fitting formulae in \citet{PP93}. For
$B=2.35\times10^{12}$~G, $\rho=0.1$ \gcc, $T=10^{5.5}$~K (the neutral
fraction $\xH\approx0.12$), the resulting absorption coefficients are
shown by the short-dashed lines in Figures \ref{fig:kkpert0} and
\ref{fig:kkpert1} for $\alpha=0$ and $+1$, respectively. For $\alpha=0$
this absorption profile is similar to the idealized model considered
above. The corresponding $\chiH_\alpha(\omega)$ are shown by the
short-dashed lines on the lower panels of Figures \ref{fig:kkpert0}
and \ref{fig:kkpert1}. In each figure, dot-dashed lines correspond to
the model of a fully ionized hydrogen plasma at the same $B$, $\rho$,
and $T$. As in the above analytic model, we see that the presence of
the atoms results in strong deviation from the fully ionized plasma
model in the vicinity of the principal atomic resonances (the
bound-free resonance for $\alpha=+1$ is almost invisible on the lower
panel of Fig.~\ref{fig:kkpert1} because of its small oscillator
strength).

\begin{figure*}\epsscale{.7}
\plotone{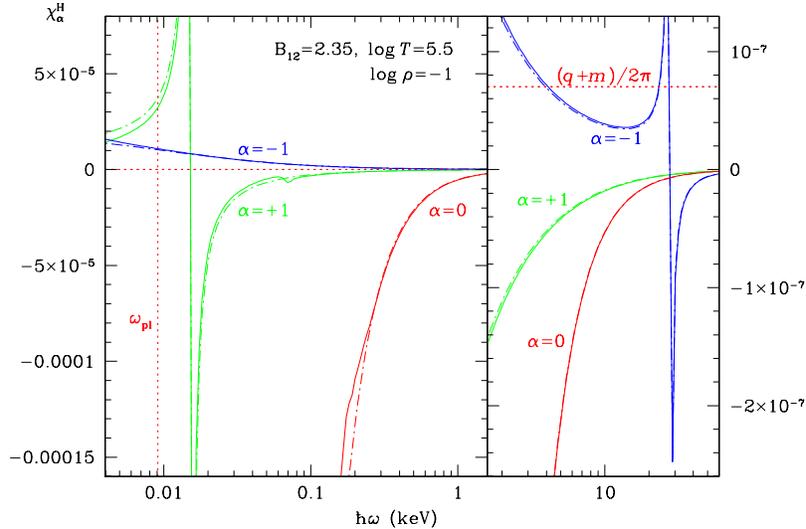}
\caption{Polarizabilities $\chiH_\alpha$ ($\alpha=\pm1,0$)
of the partially ionized (solid lines) and fully ionized (dot-dashed lines)
plasma with the same parameters as in Figures \protect\ref{fig:kkpert0},
\protect\ref{fig:kkpert1}. The right part of the figure has an enlarged
vertical scale. The dotted lines
correspond to $\ompl$ (vertical) and $(q+m)/(2\pi)$
(horizontal), where $q$ and $m$ are
the vacuum polarization parameters.
\label{fig:kk12a3}}
\end{figure*}

The absorption coefficients obtained using the perturbation 
theory of magnetic broadening (\S\ref{sect:pert})
are shown by the long-dashed lines
on the upper panels of Figures \ref{fig:kkpert0} and \ref{fig:kkpert1}.
The long-dashed lines on the lower panels show the corresponding
polarizabilities. We see that the resonant features are smoothed down
by the magnetic broadening, and the resulting curves of $\chiH_\alpha(\omega)$
do not much differ from the fully ionized plasma model.

Still greater smoothing occurs for the accurate (nonperturbative)
functions $\mu_\alpha(\omega)$ and $\chiH_\alpha(\omega)$ 
(solid lines in the figures). 
The absorption coefficients shown on the upper panels correspond to
the opacities in Figure~9 of \citet{PC03}. The significant difference of
these absorption coefficients from the perturbational ones
arises partly from the dependence of the oscillator strengths
on the atomic velocity across the field, partly from 
transitions which were dipole-forbidden
for the atom at rest and were not taken into account 
in the perturbation approximation described above,
and partly from the nonideal plasma effect of continuum
lowering. Nevertheless, despite these differences 
in $\mu_\alpha$, the difference in $\chiH_\alpha$ is much less significant.

Figure~\ref{fig:kk12a3} shows all three
components $\chiH_\alpha$ in a wider energy range.
In this figure, as well as in the previous two, the polarizabilities in the
fully ionized plasma model, shown by the dot-dashed lines,
 are obtained from \req{KK-mu},
where the free-free contribution to $\mu_\alpha(\omega)$ includes
the frequency-dependent Coulomb logarithm $\Lambda_\alpha^\mathrm{ff}$
\citep{PC03}. Because of this frequency dependence,
these polarizabilities are not identical to the ones given by
\req{chi-element}, the difference being small at $\omega\gg\ompl$
and appreciable at $\omega\lesssim\ompl$, where both the elementary theory
and the description of absorption by binary collisions (i.e., 
neglecting the collective motion effects) 
are rather inaccurate.
The solid lines, which are obtained for the partially ionized plasma,
are fairly close to the dot-dashed lines. The only prominent
features are the proton and electron cyclotron resonances
at $\hbar\omci=0.0148$ keV and $\hbar\omc=27.21$ keV.

\begin{figure}\epsscale{1.}
\plotone{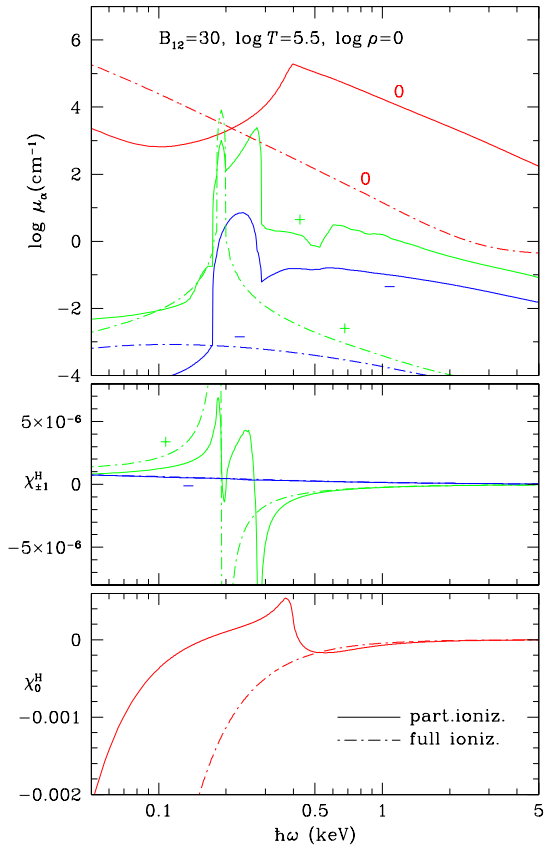}
\caption{Absorption coefficients (top panel; 
the lines marked ``+'', ``$-$'', and ``0'' correspond to
the polarization
index $\alpha=+1$, $-1$, and $0$, respectively) and
polarizabilities $\chiH_{\pm1}$ (middle) and $\chiH_0$ (bottom panel)
of the partially ionized (solid lines) and fully ionized (dot-dashed lines)
plasma for $B=3\times10^{13}$~G, $\rho=1$ \gcc, and $T=3.16\times10^5$~K.
\label{fig:kk13a4}}
\end{figure}

However, such smoothing does not always occur. Let us consider a higher 
magnetic field $B=3\times10^{13}$~G and density $\rho=1$ \gcc,
retaining the same temperature. Then $\xH=0.89$.
Figure~\ref{fig:kk13a4} demonstrates
the absorption coefficients and polarizabilities for the
fully ionized (dot-dashed lines) and partially ionized
(solid lines) plasma. In addition to the proton cyclotron 
resonance at $\hbar\omega=0.19$ keV,
absorption coefficients in the partially ionized plasma
exhibit magnetically broadened bound-bound
and bound-free features. The most prominent are the bound-bound
absorption at $\hbar\omega\approx0.2$--0.3 keV for $\mu_{+1}$
and the photoionization edge at $\hbar\omega=0.408$ keV for $\mu_0$.
These features are clearly reflected in the behavior of $\chiH_{+1}$
and $\chiH_0$, shown in the lower panels. Thus, with increasing $B$,
the abundance of neutral states increases along with their influence
on the polarizability.

A similar trend occurs with lowering $T$. If, for example,
in the previous case ($B=2.35\times10^{12}$~G, $\rho=0.1$ \gcc)
we take a lower $T=1.58\times10^5$~K, then we will have 
94.1\% of H atoms in the ground state, 1.3\% in the excited states,
and 1.1\% of H$_2$ molecules. At $\rho=1$ \gcc,
the abundance of the atoms will be 96.9\% (with roughly equal
fractions of excited states and molecules), and we
nearly recover the case studied by
\citet{BulikPavlov}, who assumed 100\% of 
atoms for these plasma parameters.

\section{Effect of Partial Ionization on Polarization and Opacities 
of Normal Modes}
\label{sect:res}

In the coordinate system $xyz$ with $z$ along the wave vector 
of the photon
and $\bm{B}$ in the $x$--$z$ plane,
the polarization vectors $\bm{e}^j$
of the normal modes in a magnetized plasma
can be written as \citep{HoLai02}
\beq
  (e^j_x,e^j_y,e^j_z)=(1+K_j^2+K_{z,j}^2)^{-1/2} \,
  (i K_j, 1, i K_{z,j}),
\eeq
where $j=1,2$ correspond to the extraordinary mode (X-mode)
and ordinary mode (O-mode).
The parameters $K_j$ and $K_{z,j}$ are expressed in terms 
of the components of the dielectric and magnetic tensors as
\bea
&&\hspace*{-2em}
  K_j = \beta \left\{
   1 + (-1)^j \left[ 1 + \frac{1}{\beta^2} 
   + \frac{m}{1+\hat{a}} \frac{\sin^2\theta_B}{\beta^2}\right]^{1/2}
   \right\},
\\&&
  K_{z,j} = - \frac{ 
   (\varepsilon' - \eta') K_j \cos\theta_B + g
   }{
   \varepsilon' \sin^2\theta_B + \eta' \cos^2\theta_B } 
   \, \sin\theta_B,
\eea
where
\beq
 \beta = \frac{\eta' - \varepsilon' + g^2/\varepsilon' + \eta' \,m/(1+\hat{a})
   }{
   2 \, g }
   \, \frac{ \varepsilon'}{\eta'}
   \,\frac{\sin^2\theta_B}{\cos\theta_B},
\eeq
$\theta_B$ is the angle between $\bm{B}$ and the $z$ axis,
and (see Eqs.~[\ref{linear}]--[\ref{chivac}]) 
$\varepsilon' = \varepsilon + \hat{a}$, $\eta' = \eta + \hat{a} + q$.
In the usual case where the plasma and vacuum polarizabilities are small
($|\chiH_\alpha| \ll (4\pi)^{-1}$ and $|\hat{a}|,q,|m| \ll 1$),
the polarization parameter $\beta$ is given by
\beq
  \beta \approx \frac{2\chiH_0 - \chiH_{+1} - \chiH_{-1} + (q+m)/(2\pi)
  }{
  2\,(\chiH_{+1} - \chiH_{-1}) }\,\frac{\sin^2\theta_B}{\cos\theta_B} .
\label{beta-approx}
\eeq

The opacity in the mode $j$ can be written as
\beq
  \kappa_j(\omega,\theta_B) = \sum_{\alpha=-1}^1 \!\!
   |e_\alpha^j(\omega,\theta_B)|^2 \,\hat\kappa_\alpha(\omega),
\label{opac}
\eeq
where $\hat\kappa_\alpha$ ($\alpha=-1,0,1$) do not depend on
$\theta_B$. For a partially ionized, strongly magnetized
hydrogen plasma $\hat\kappa_\alpha(\omega)$ have been
obtained by \citet{PC03,PC04}.

The polarization vectors $\eX$ and $\eO$ for the
polarizabilities presented in Figure~\ref{fig:kk12a3} prove to be
indistinguishable from the results for the fully ionized plasma
at general $\theta_B$ values. They
exhibit vacuum polarization resonances at $\hbar\omega \sim 10$ keV
related to intersections of $\chiH_{-1}$ with the combination of vacuum
polarization coefficients $(q+m)/2\pi$ that enters \req{beta-approx}
(the horizontal dotted line in Fig.~\ref{fig:kk12a3}; in this case
$|\chiH_0|,|\chiH_{+1}|\ll|\chiH_{-1}|$) and the electron cyclotron
resonance at $\hbar\omega\approx27$ keV. If we neglected the CM
coupling effect, we would observe additional resonant features near
$\hbar\omega=0.07$ and 0.2 keV, associated with the bound-bound
transitions for $\alpha=+1$ (Fig.~\ref{fig:kkpert1}) and $\alpha=0$
(Fig.~\ref{fig:kkpert0}). 
Actually these features are smoothed away by the CM coupling.

\begin{figure}\epsscale{1.}
\plotone{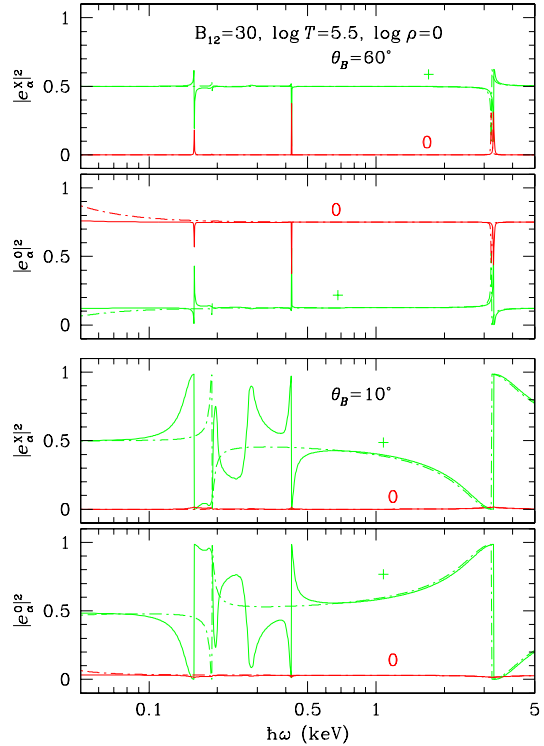}
\caption{Squared moduli of the cyclic components $e^j_\alpha$ of
polarization vectors $\eX$ and $\eO$
for the normal waves propagating at angles $\theta_B=60^\circ$
(two upper panels) and $10^\circ$ (two lower panels)
in a hydrogen plasma
with $B=3\times10^{13}$~G, $T=3.16\times10^5$~K,
$\rho=1$ \gcc.
 Solid lines: partially ionized plasma; dot-dashed lines: full ionization.
\label{fig:kkm13a4}}
\end{figure}

\begin{figure*}\epsscale{.8}
\plotone{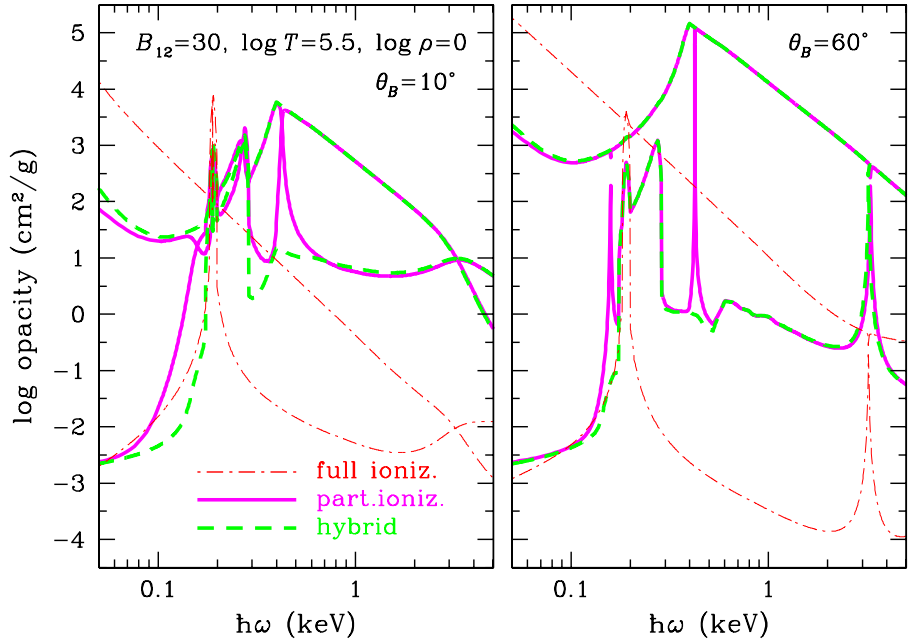}
\caption{Opacities in two modes $\kappa_j$ (Eq.~[\ref{opac}])
for the same plasma parameters as in Fig.~\protect\ref{fig:kkm13a4}
and two angles, $\theta_B=10^\circ$ (left panel)
and $\theta_B=60^\circ$ (right panel). 
Solid lines: a self-consistent calculation
for a partially ionized hydrogen plasma;
dashed lines: partially ionized hydrogen plasma within a ``hybrid''
treatment
(a model that uses the basic opacities $\hat\kappa_\alpha$
from the model of partially ionized plasma and polarization vectors
$\bm{e}^j$
from the model of full ionization);
light dot-dashed lines: model of full ionization.
The lower curve of each type is related to extraordinary and the upper
one to ordinary mode (the dashed curve often coincides 
with the solid one).
\label{fig:ang13a4}}
\end{figure*}

\begin{figure*}\epsscale{.8}
\plotone{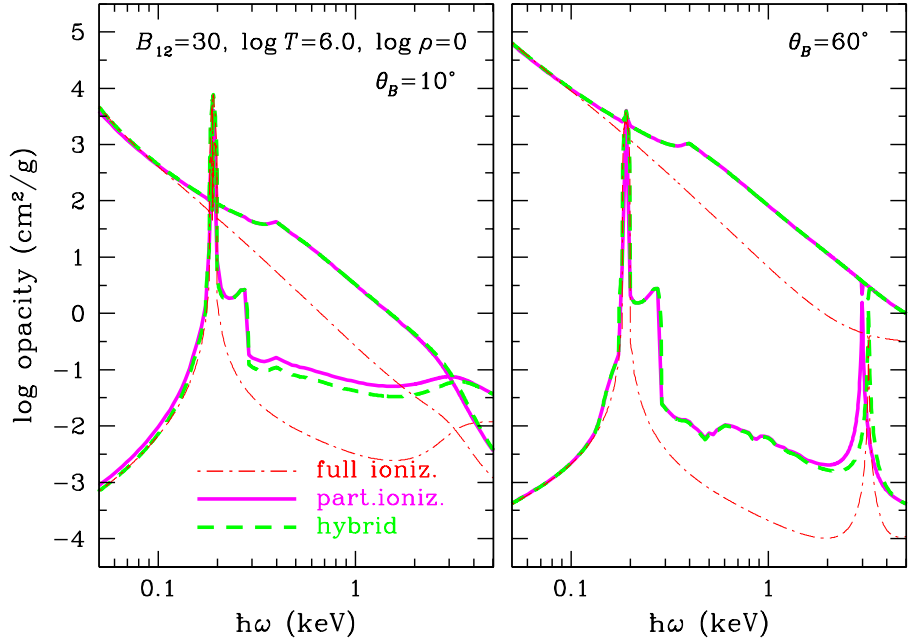}
\caption{Same as in Fig.~\protect\ref{fig:ang13a4}
but for higher temperature, $T=10^6$~K.
\label{fig:ang13a4b}}
\end{figure*}

For the second set of plasma parameters considered in \S\ref{sect:chi-real},
the ionization degree is relatively low ($1-\xH = 0.11$).
Figure~\ref{fig:kkm13a4} shows squared moduli of two components
of the polarization vectors, $e^j_0$ and $e^j_{+1}$
(the third component can be found as $|e_{-1}|^2=1-|e_{+1}|^2-|e_0|^2$)
for two propagation angles $\theta_B=60^\circ$ and $\theta_B=10^\circ$.
Dot-dashed and solid curves correspond to the fully and partially ionized
plasma models, respectively. At $\theta_B=60^\circ$ (two upper panels), 
there are two sharp
resonances for the partially ionized case at $\hbar\omega=0.158$
and 0.425 keV, associated with the zero level crossings
by $\chiH_0(\omega)$ (the bottom panel of Fig.~\ref{fig:kk13a4}), which are 
absent in the case of full ionization. The feature at 3.3 keV
is the vacuum resonance.
At smaller angle (two lower panels of Fig.~\ref{fig:kkm13a4}), 
these resonances become broader,
and there appear additionally other features associated with the behavior
of $\chiH_{+1}$ (see the middle panel of Fig.~\ref{fig:kk13a4}).
For the fully ionized plasma, an additional feature is
just the proton cyclotron resonance, whereas for the partially ionized
case the behavior of the polarization vectors is more complicated
because of the influence of the atomic resonances.

Figure~\ref{fig:ang13a4} shows the opacities calculated
according to \req{opac}, using $|e^j_\alpha|^2$ 
shown in Figure~\ref{fig:kkm13a4}
and $\hat\kappa_\alpha(\omega)$
calculated as in \citet{PC03,PC04}.
The opacities that take into account partial ionization
are plotted by solid lines, and those assuming full ionization by
dot-dashed lines. The dashed lines correspond to a hybrid approach,
where the polarization
of normal modes is described by
the formulae for a fully ionized plasma,
and
$\hat\kappa_\alpha(\omega)$ take into account bound-bound 
and bound-free atomic transitions
\citep{PC03}. Although this approach
is closer to reality than the model of full ionization,
there are significant differences from the more accurate result
drawn by the solid lines. In particular, the feature near 
$\hbar\omega=0.5$ keV is missed in the hybrid approximation.

With increasing temperature, the differences between
the self-consistent and hybrid approximations go away.
Figure~\ref{fig:ang13a4b} shows the case where the plasma
parameters are the same as in Figure~\ref{fig:ang13a4},
except for $T=10^6$~K. At this temperature, $\xH=1.4$\%.
Such small amount of neutral atoms is still very important 
for the opacities, but the hybrid approximation
yields the result close to self-consistent one.

Summarizing, we conclude that the hybrid approach to
calculation of the mode opacities can be applicable
if the fraction of bound states in the plasma, $\xH$
is small.

\begin{figure}\epsscale{1.}
\plotone{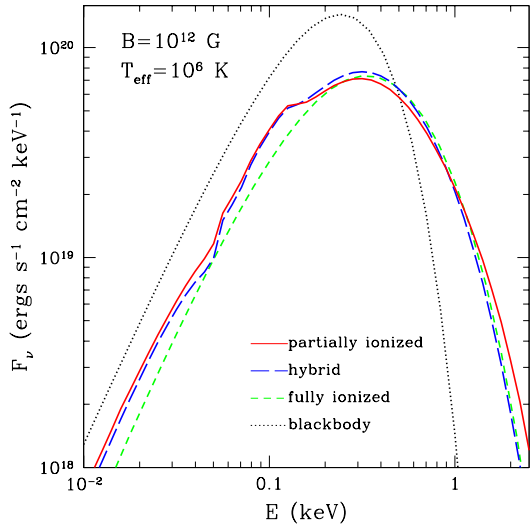}
\caption{Spectrum of hydrogen atmosphere with $B=10^{12}$~G
(field normal to the surface)
and effective temperature
$\Teff=10^6$~K. The solid line is for the self-consistent
model of a partially ionized atmosphere, the long-dashed line 
is for the hybrid model, the short-dashed line is for a fully
ionized atmosphere, and the dotted line is for a blackbody
with $T=\Teff$.
\label{fig:f121}}
\end{figure}

\section{Spectra}
\label{sect:spectra}

\begin{figure}\epsscale{1.}
\plotone{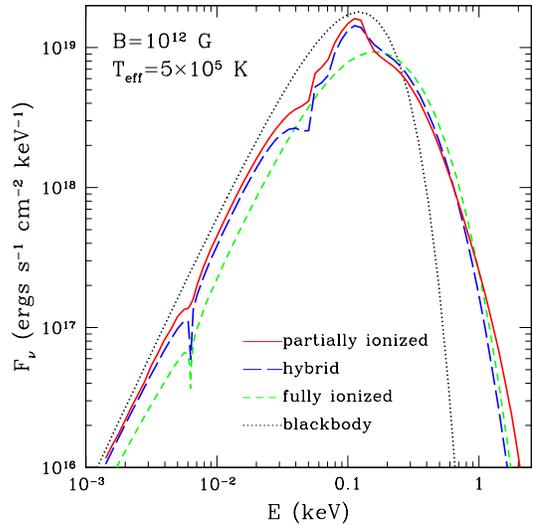}
\caption{Same as in Fig.~\protect\ref{fig:f121}
but for lower effective temperature, $\Teff=5\times10^5$~K.
\label{fig:f1255}}
\end{figure}

\begin{figure}\epsscale{1.}
\plotone{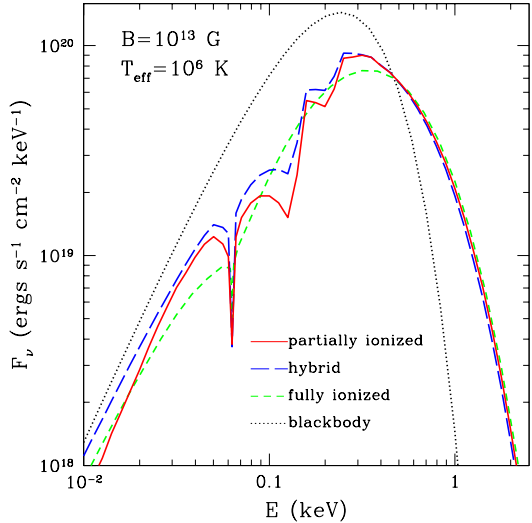}
\caption{Same as in Fig.~\protect\ref{fig:f121}
but for higher field strength, $B=10^{13}$~G.
\label{fig:f131}}
\end{figure}

Examples of application of the self-consistent opacity calculation 
for partially ionized hydrogen NS atmospheres are given in Figures
\ref{fig:f121}--\ref{fig:f131}. Here, the atmosphere parameters are the
same as for the low-field models in \citet{hoetal03}. The solid lines
are obtained using the self-consistent calculation of the opacities,
while the long-dashed lines reproduce the hybrid treatment described above.
The fully ionized plasma model (short dashes) and blackbody (dots) are
shown for comparison. The difference in the spectra obtained using the
self-consistent and hybrid approaches is partly due to the difference
in the temperature profiles within the atmosphere. As
expected, this difference is small at relatively low magnetic field
$B=10^{12}$~G and effective temperature $\Teff=10^6$~K
(Fig.~\ref{fig:f121}), where the fraction of atoms is small at every
optical depth in the atmosphere, but it becomes larger for lower
temperature ($\Teff=5\times10^5$~K, Fig.~\ref{fig:f1255}) or stronger
field ($B=10^{13}$~G, Fig.~\ref{fig:f131}). In the case shown in 
Figure~\ref{fig:f1255}, the temperature profile $T(\rho)$ 
calculated with the new
(self-consistent) opacities is higher by $\sim10$\%, so that photons
come from shallower (lower atomic fraction) layers of the atmosphere, 
which results in weaker spectral features due to the atomic
transitions. A modification of the temperature profile is also
responsible for the weaker proton cyclotron feature in this case. At
contrast, in the case shown in Figure~\ref{fig:f131} the temperature
profile is lower by $\lesssim10$\%, and photons come from deeper (higher
atomic fraction) layers, resulting in stronger atomic features.

For superstrong fields ($B\gtrsim10^{14}$~G), the current atmosphere
models are less reliable because of the unsolved problems of mode
conversion and dense plasma effect \citep[e.g.,][]{hoetal03}, whose
importance increases with increasing $B$.

\section{Conclusions and Outlook}
\label{sect:concl}

We have studied the polarizability and electromagnetic polarization
modes in a partially ionized, strongly magnetized hydrogen plasma. The
full account of the coupling of the quantum mechanical structure of the
atoms to their center-of-mass motion across the magnetic field is shown
to be crucial for the correct evaluation of the polarization properties
and opacities of the plasma. The self-consistent treatment of the
polarizability and absorption coefficients is ensured by use of the
Kramers-Kronig relation. Such treatment proves to be important if the
ionization fraction of the plasma is low ($\lesssim50$\%). For high
degree of ionization ($\gtrsim 80$\%), the polarizability of a fully
ionized plasma remains a good approximation, just as previously assumed
\citep{PC03}. This approximation was adopted in the NS atmosphere
models built in \citet{hoetal03,Ho-COSPAR}. A comparison with updated
spectra based on the self-consistent treatment (\S\ref{sect:spectra})
shows that this approximation is satisfactory if $B\lesssim10^{13}$~G
and $\Teff\gtrsim10^6$~K. The self-consistent treatment is
needed in the atmospheres of cool or ultramagnetized NSs, 
with relatively low degrees of
ionization.

There are several limitations of the present model,
which may become important for the magnetar fields
and/or for low $\Teff$. 
While H atoms are treated accurately in our calculations of the 
EOS and opacities, H$_2$ molecules are included in the EOS 
using the static approximation (i.e., without their CM coupling)
and neglected in the opacities. Other bound species, such as
H$_2$$^+$ \citep[e.g.,][]{TurbLop03}, 
H$_3$$^{2+}$ \citep{LopTurb}, and
H$_n$ chains \citep*{LSS92} are not included.
Moreover, the NS may have a condensed surface, 
with negligible vapor above it
(\citealt{LS97}; \citealt{Lai01}).
For a NS with mass $M=1.4~\Msun$ and radius $R=10$ km,
estimates
at $\log B$ (G)$=13.5$--15
\citep[based on][]{PC04}
indicate that a thick H atmosphere will be present, and a condensed
surface will not occur, provided
$\Teff \gtrsim 3.8\times10^5\,(B/10^{14}\mbox{ G})^{1/4}$~K;
slightly higher $\Teff$ is needed to ensure
negligible abundance of molecules. 

Another uncertainty in ultramagnetized NS atmospheres
is the dense plasma effect: the decoupling layer for photons 
in the atmosphere (where optical depth $\approx1$) may occur
at high density where the electron plasma frequency
exceeds the photon frequency \citep[e.g.,][]{hoetal03,Lloyd03}.
The present treatment is not applicable for such cases.
In addition,
construction of reliable atmosphere models 
at $B\gtrsim 10^{14}$~G
requires solution
of the problem of mode conversion \citep{LaiHo03a}.

Furthermore, for fitting observed spectra one should 
construct a grid of models with different field orientations
and a range of field strengths,
and produce angle- and field-integrated synthetic spectra
for an assumed field geometry. Since all
the discussed spectral resonances are
$B$-dependent, and some of them are $\theta_B$-dependent,
we expect that such integration will somewhat smooth
the spectral features \citep{HoLai04}.


\acknowledgements
A.P.\ acknowledges the hospitality
of the Astronomy Department of Cornell University
and the theoretical astrophysics group 
at the Ecole Normale Sup\'erieure de Lyon.
The work of A.P.\ is supported in part by RFBR grants 
02-02-17668 and 03-07-90200, 
and ``Russian Leading Scientific Schools''
grant 1115.2003.2. 
D.L.\ is supported in part by NSF grant AST 0307252,
NASA grant NAG 5-12034,
and SAO grant TM4-5002X.
W.H.\ is supported by NASA through Hubble Fellowship grant
HF-01161.01-A awarded by STScI, which is operated by AURA, Inc.,
for NASA, under contract NAS 5-26555.

\appendix
\section{Fitting Formulae for the Vacuum Polarization Coefficients}
The studies of the vacuum polarization by strong fields have long
history; an extensive bibliography is given by \citet{Schubert}. In the
low-energy limit, $\hbar\omega\ll\mel c^2$, 
the Euler-Heisenberg effective Lagrangian can be
applied. The relevant dimensionless magnetic-field parameter is
 $b\equiv B/\Bq$, where $\Bq=\mel^2c^3/e\hbar=4.414\times 10^{13}$~G. 
In the limit 
$b\ll 1$, the vacuum polarization coefficients are given by 
\citep{Adler}
\beq
 \hat{a}= - \frac{2 \alphaf}{45\pi}\, b^2,
\quad
 q= \frac{7 \alphaf}{45\pi}\, b^2,
\quad
 m= - \frac{4 \alphaf}{45\pi}\, b^2,
\label{vac-weak}
\eeq
where $\alphaf=e^2/\hbar c = 1/137.036$ is the fine-structure constant.
For arbitrary $B$, the tensors of vacuum polarization
 $\bm{\chi}^\mathrm{vac}$ and $\bm{\mu}$
have been obtained by \citet{hehe} in terms of special functions
and by \citet{kohriyam} numerically. The result of \citet{hehe}
can be reduced to the following more convenient form:
\bea
&&
 \hat{a} = \frac{\alphaf}{2\pi} \bigg[
 \xi X(\xi) - 2 \int_1^\xi X(\xi')\, \dd \xi' - 0.0329199 \,\bigg],
\label{hehe-a}
\\&&
 \hat{a} + q = \frac{\alphaf}{2\pi} \bigg[\,
 \frac{2}{9\xi^2} -\frac23 X'(\xi) \,\bigg],
\label{hehe-q}
\\&&
 m = \frac{\alphaf}{2\pi} \Big[\,
 \xi X(\xi) - \xi^2 X'(\xi) \Big],
\label{hehe-m}
\eea
where $X(\xi)$ is expressed through the Gamma function $\Gamma(x)$:
\bea
 X(\xi) &=& 2 \ln \Gamma(\xi/2) - \frac{1}{3\xi}
 - \ln \frac{4\pi}{\xi} 
 + \xi + \xi \ln \frac{2}{\xi},
\\&&
 X'(\xi) = \dd X(\xi)/\dd\xi,
 \qquad 
 \xi = b^{-1}.
\label{hehe-xi}
\eea
Results of
calculation according to Eqs.~(\ref{hehe-a})--(\ref{hehe-xi})
agree with Fig.~2 of \citet{kohriyam} and can be
approximately represented by 
\bea
&&
 \hat{a} \approx - \frac{2\alphaf}{9\pi} \ln\bigg(
 1 + \frac{b^2}{5}\,\frac{
  1+0.25487\,b^{3/4}
  }{
  1+0.75\,b^{5/4}}\bigg),
\label{fit-a}
\\&&
 q \approx \frac{7\alphaf}{45\pi}\,b^2\,\frac{
 1 + 1.2\,b
 }{
 1 + 1.33\,b + 0.56\,b^2
 },
\label{fit-q}
\\&&
 m \approx - \frac{\alphaf}{3\pi} \, \frac{ b^2 }{
 3.75 + 2.7\,b^{5/4} + b^2 }.
\label{fit-m}
\eea
Equations (\ref{fit-a})--(\ref{fit-m}) exactly recover the weak-field
limits (\ref{vac-weak}) and the leading terms in 
the high-field ($b\gg1$) expansions 
(Eqs.\ [2.15]--[2.17] of \citealt{HoLai02});
in the latter regime, 
Eqs.\ (\ref{fit-a}) and (\ref{fit-q}) ensure also the 
terms next to leading. The maximum errors
are 1.1\% at $b=0.07$ for \req{fit-a}, 2.3\% at $b=0.4$ for \req{fit-q},
and 4.2\% at $b=0.3$ for \req{fit-m}.


\end{document}